\documentclass[runningheads]{llncs}
  \usepackage{amsfonts}
  \usepackage{amsmath}
  \usepackage{color}
  \usepackage{graphicx}
  \usepackage{tikz}
  \usepackage{listings}
  \usepackage{color}
  \usepackage{hyperref}
  
  \usepackage{subcaption}
  \usetikzlibrary{positioning}
  \usetikzlibrary{shapes}
  \usetikzlibrary{shapes.geometric}
  \usetikzlibrary{arrows}
  \usetikzlibrary{graphs}
  
  \definecolor{eclipseBlue}{RGB}{42,0.0,255}
  \definecolor{eclipseGreen}{RGB}{63,127,95}
  
  \newcommand{\todoJ}[1]{}
  \newcommand{\todoY}[1]{}
  \newcommand{\pgt}[1]{\texttt{PGT}}
  \newcommand{\psl}[1]{\texttt{PSL}}
  \newcommand{\fancy}[1]{\ensuremath{\mathcal{#1}}}

\begin{document}
  \title{Goal-Oriented Conjecturing for Isabelle/HOL}
  
  %
  %
  \author{Yutaka Nagashima\inst{1,2}\thanks{Supported by the European Regional Development Fund under the project AI \& Reasoning (reg. no.CZ.02.1.01/0.0/0.0/15\_003/0000466)} \and 
  Julian Parsert\inst{2}\thanks{Supported by the European Research Council (ERC) grant no 714034 SMART.}\\
  \email{\{yutaka.nagashima, julian.parsert\}@uibk.ac.at}}
  \authorrunning{Yutaka Nagashima \and Julian Parsert}
  %
  \institute{
   CIIRC, Czech Technical University in Prague, Czech Republic
   \and
   Department of Computer Science, 
   University of Innsbruck, Austria
  }
  %
  \maketitle 
  \begin{abstract}
    We present \pgt{},
    a \underline{P}roof \underline{G}oal \underline{T}ransformer for Isabelle/HOL.
    Given a proof goal and its background context,
    \pgt{} attempts to generate conjectures from the original goal
    by transforming the original proof goal.
    These conjectures should be weak enough to be provable by automation
    but
    sufficiently strong to prove the original goal.
    By incorporating \pgt{} into the pre-existing \psl{} framework, 
    we exploit Isabelle's strong automation to identify and prove such conjectures.
  \end{abstract}
  
  \section{Introduction} \label{sec:introduction}
  Consider the following two reverse functions 
  defined in literature~\cite{tutorial}:
  \begin{verbatim}
  primrec itrev:: "'a list ⇒ 'a list ⇒ 'a list" where
   "itrev [] ys = ys" | "itrev (x#xs) ys = itrev xs (x#ys)"
  primrec rev :: "'a list ⇒ 'a list" where
   "rev [] = []" | "rev (x # xs) = rev xs @ [x]"
  \end{verbatim}
  \noindent
  How would you prove their equivalence \texttt{"itrev xs [] = rev xs"}?
  Induction comes to mind. 
  However, it turns out that Isabelle's default proof methods,
  \verb|induct| and \verb|induct_tac|, are unable to handle this proof goal
  effectively.
  
  Previously, we developed \psl{} \cite{psl},
  a programmable, meta-tool framework for Isabelle/HOL.
  With \psl{} one can write the following strategy for induction:
  \begin{verbatim}
  strategy DInd = Thens [Dynamic (Induct), Auto, IsSolved]
  \end{verbatim}
  \psl{}'s \verb|Dynamic| keyword creates
  variations of the \verb|induct| method
  by specifying different combinations of promising arguments
  found in the proof goal and its background proof context.
  Then, \verb|DInd|
  combines these induction methods with the general purpose proof method, \verb|auto|,
  and \verb|is_solved|,
  which checks if there is any proof goal left after applying \verb|auto|.
  As shown in Fig. \ref{fig:psl},
  \psl{} keeps applying the combination of
  a specialization of \verb|induct| method and \verb|auto|,
  until either \verb|auto| discharges all remaining sub-goals or
  \verb|DInd| runs out of the variations of \verb|induct| methods
  as shown in Fig.~\ref{fig:psl}.
  
  This approach works well only if the resulting sub-goals 
  after applying some \verb|induct| are
  easy enough for Isabelle's automated tools (such as \verb|auto| in \verb|DInd|) to prove.
  When proof goals are presented in an automation-unfriendly way,
  however, it is not enough to set a certain combination of arguments
  to the \verb|induct| method.
  In such cases engineers have to investigate the original goal
  and come up with auxiliary lemmas,
  from which they can derive the original goal.
  
  In this paper, we present \pgt{},
  a novel design and prototype implementation\footnote{available at Github \href{https://github.com/data61/PSL/releases/tag/v0.1.1}{https://github.com/data61/PSL/releases/tag/v0.1.1}.
  The example of this paper appears in \texttt{PSL/PGT/Example.thy}.} of
  a conjecturing tool for Isabelle/HOL.
  We provide \pgt{} as an extension to \psl{}
  to facilitate the seamless integration with other Isabelle sub-tools.
  Given a proof goal,
  \pgt{} produces a series of conjectures
  that might be useful in discharging the original goal,
  and \psl{} attempts to identify the right one
  while searching for a proof of
  the original goal using those conjectures.
  \section{System Description}
  \subsection{Identifying Valuable Conjectures via Proof Search}\label{sec:overview}
  To automate conjecturing,
  we added the new language primitive, \verb|Conjecture| to \psl{}.
  Given a proof goal,
  \verb|Conjecture| first produces a series of conjectures that might be useful
  in proving the original theorem,
  following the process described in Section \ref{sec:conjecturing}.
  For each conjecture, \pgt{} creates a \verb|subgoal_tac| method
  and inserts the conjecture as the premise of the original goal.
  When applied to \texttt{"itrev xs [] = rev xs"}, for example,
  \verb|Conjecture| generates the following proof method along with 130 other variations of the \verb|subgoal_tac| method:
  \begin{verbatim}
  apply (subgoal_tac "!!Nil. itrev xs Nil = rev xs @ Nil")
  \end{verbatim}
  \noindent
  where \verb|!!| stands for the universal quantifier in Isabelle's meta-logic.
  Namely, \verb|Conjecture| introduced a variable of name \verb|Nil| 
  for the constant \verb|[]|.
  Applying this method to the goal results in the following two new sub-goals:
  \begin{verbatim}
  1. (!!Nil. itrev xs Nil = rev xs @ Nil) ==> itrev xs [] = rev xs
  2. !!Nil. itrev xs Nil = rev xs @ Nil
  \end{verbatim}
  \verb|Conjecture| alone cannot determine
  which conjecture is useful for the original goal.
  In fact, some of the generated statements are not even true or provable.
  To discard these non-theorems and to reduce the size of \psl{}'s search space,
  we combine \verb|Conjecture|
  with \verb|Fastforce| (corresponding to the \verb|fastforce| method)
  and \verb|Quickcheck| (corresponding to Isabelle's sub-tool \textit{quickcheck}~\cite{quickcheck}) sequentially
  as well as \verb|DInd| as follows:
  \begin{verbatim}
  strategy CDInd = Thens [Conjecture, Fastforce, Quickcheck, DInd]
  \end{verbatim}
  \noindent
  Importantly, \verb|fastforce| does not return an intermediate proof goal:
  it either discharges the first sub-goal completely or
  fails by returning an empty sequence.
  Therefore, whenever \verb|fastforce| returns a new proof goal
  to a sub-goal resulting from \verb|subgoal_tac|,
  it guarantees that the conjecture inserted
  as a premise is strong enough for Isabelle to prove the original goal.
  In our example, the application of \verb|fastforce|
  to the aforementioned first sub-goal succeeds,
  changing the remaining sub-goals to the following:
  \begin{verbatim}
  1. !!Nil. itrev xs Nil = rev xs @ Nil
  \end{verbatim}
  However, \psl{} still has to deal with many non-theorems:
  non-theorems are often strong enough to imply the original goal
  due to the principle of explosion.
  Therefore,
  \verb|CDInd| applies \verb|Quickcheck| to discard
  easily refutable non-theorems.
  The atomic strategy \verb|Quickcheck| returns the same sub-goal
  only if Isabelle's sub-tool quickcheck
  does not find a counter example,
  but returns an empty sequence otherwise.
  
\begin{figure}[t]
      \centering
      \begin{subfigure}[]{0.48\textwidth}
          \includegraphics[width=\textwidth]{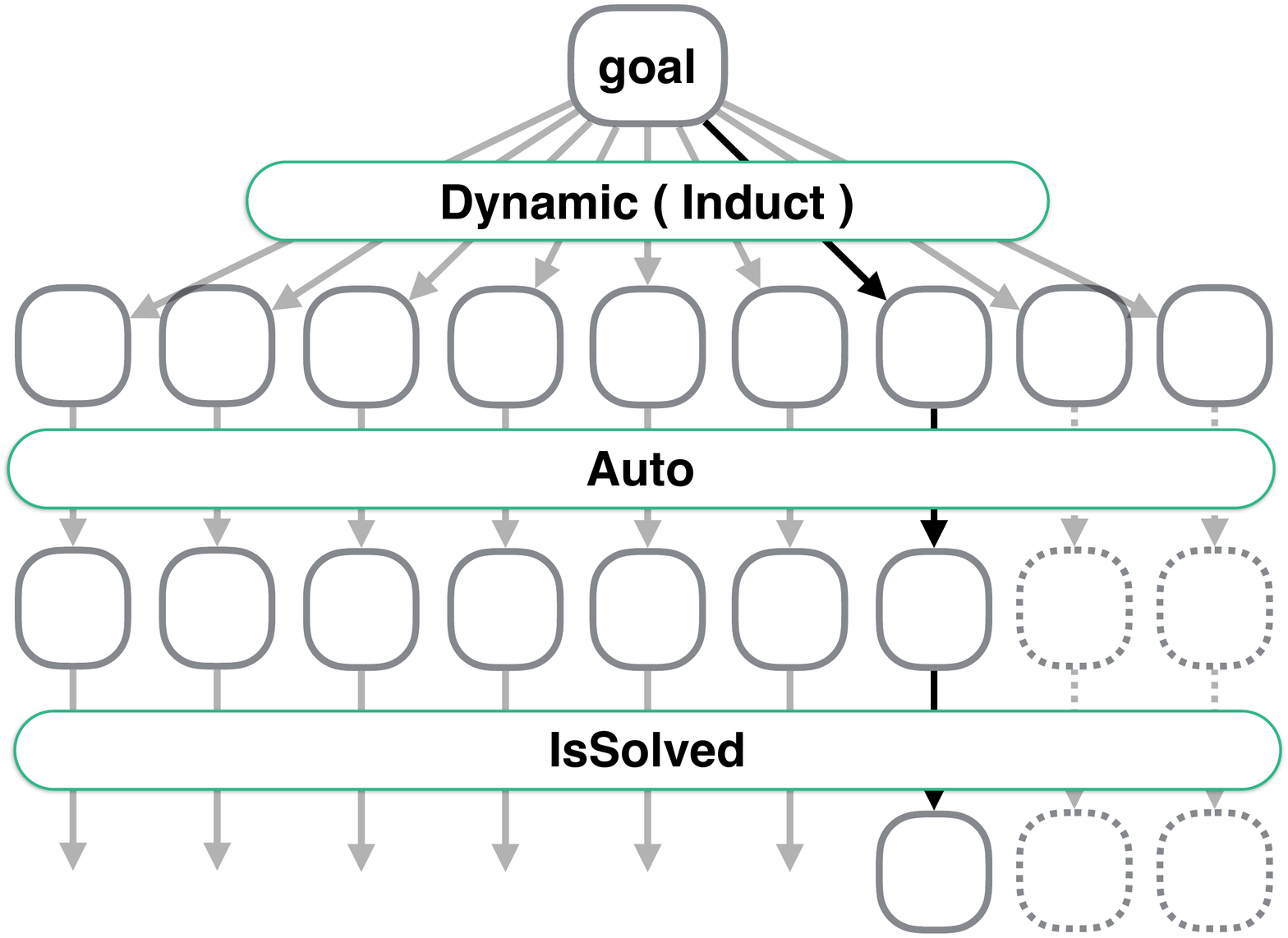}
          \caption{Search tree of \texttt{DInd}}
          \label{fig:psl}
      \end{subfigure}
      ~ 
      \begin{subfigure}[]{0.48\textwidth}
          \includegraphics[width=\textwidth]{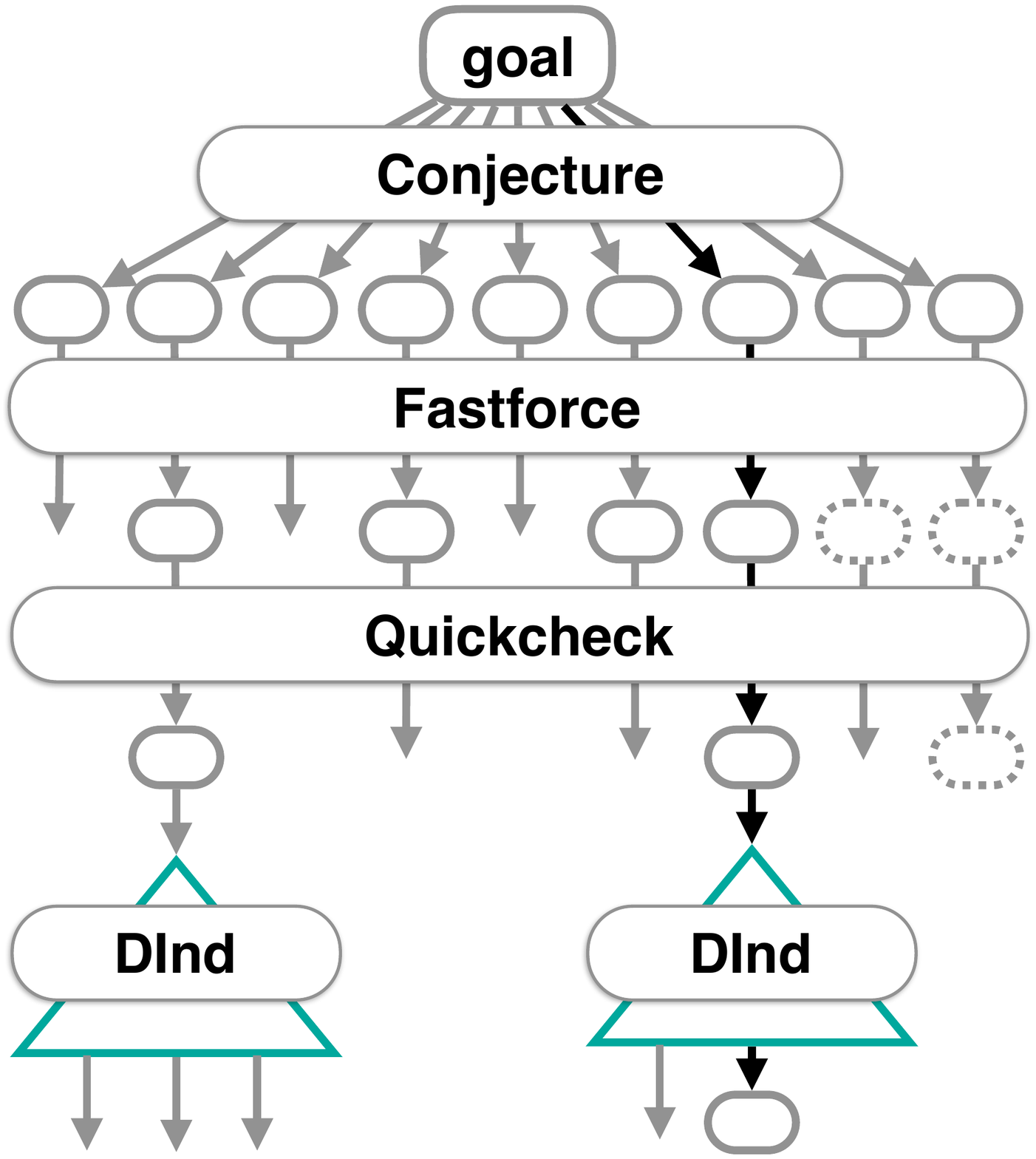}
          \caption{Search tree of \texttt{CDInd}}
          \label{fig:pgt}
      \end{subfigure}
      \caption{\psl{}'s proof search with/without \pgt{}.}
      \label{fig:example_search_tree}
  \end{figure}

  Now we know that
  the remaining conjectured goals are 
  strong enough to imply the original goal and
  that they are not easily refutable.
  Therefore, \verb|CDInd| applies its sub-strategy \verb|DInd|
  to the remaining sub-goals and it stops its proof search
  as soon as it finds the following proof script, which will be printed in Isabelle/jEdit's output panel.
  \begin{verbatim}
    apply (subgoal_tac "!!Nil. itrev xs Nil = rev xs @ Nil")
    apply fastforce apply (induct xs) apply auto done
  \end{verbatim}
  Fig. \ref{fig:pgt} shows how \verb|CDInd|
  narrows its search space 
  in a top-down manner.
  Note that \psl{} lets you use other Isabelle sub-tools to prune conjectures.
  For example, you can use both \textit{nitpick} \cite{nitpick} 
  and quickcheck: \texttt{Thens [Quickcheck, Nitpick]} in \texttt{CDInd}.
  It also let you combine \verb|DInd| and \verb|CDInd| into one:
  \texttt{Ors [DInd, CDInd]}.
  
  \subsection{Conjecturing}\label{sec:conjecturing}
\begin{figure}[t]
  \centering
  \scalebox{0.5}{
  \begin{tikzpicture}[%
      ->,
      shorten >=2pt,
      >=stealth,
      node distance=0.4cm,
      noname/.style={%
        ellipse,
        minimum width=5em,
        minimum height=3em,
        draw
      }]
      \tikzstyle{every node}=[font=\LARGE]
      \node[rectangle,draw] (2)
        {Extract constants and common sub-terms from the original goal \fancy{T}};
      \node[rectangle,draw] (3) [below=of 2]
        {Generalize \fancy{T} to produce $\fancy{C}_0,\dots,\fancy{C}_n$};
      \node[rectangle,draw] (4) [below=of 3]
        {Call \texttt{conjecture} for goal oriented conjecturing~(Fig.~\ref{fig-simpConj}) for each \fancy{T} and $\fancy{C}_0,\dots,\fancy{C}_n$};
      \node[rectangle,draw] (5) [below=of 4]              
        {Clean \& return};
      \path (2) edge                   node {} (3)
            (3) edge                   node {} (4)
            (4) edge                   node {} (5);
  \end{tikzpicture}
  }
  \caption{The overall workflow of \texttt{Conjecture}.}
  \label{fig:conjecturing}
  \end{figure}
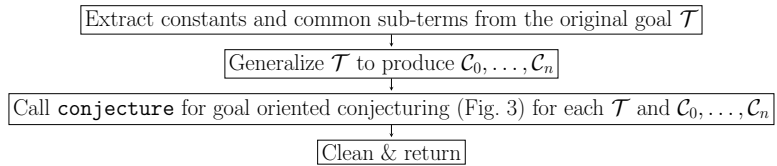

  Section \ref{sec:overview} has described
  how we identify useful conjectures.
  Now, we will focus on how \pgt{} creates conjectures in the first place.
  \pgt{} introduced both automatic conjecturing (\verb|Conjecture|) and
  automatic generalization (\verb|Generalize|).
  Since the conjecturing functionality uses generalization,
  we will only describe the former.
  We now walk through the main steps
  that lead from a user defined goal to a set of potentially \emph{useful} conjectures, as illustrated in Fig. \ref{fig:conjecturing}.
  %
  We start with the extraction of constants and sub-terms, 
  continue with generalization, goal oriented conjecturing, 
  and finally describe how the resulting terms are sanitized.
  
  \paragraph{Extraction of Constants and Common Sub-terms.}
  Given a term representation \fancy{T} of the original goal, 
  \pgt{} extracts the constants and sub-terms that appear multiple times in
  \fancy{T}.
  In the example from Section~\ref{sec:introduction},
  \pgt{} collects the constants
  \verb|rev|, \verb|itrev|, and \verb|[]|.
  \paragraph{Generalization.} 
  Now, \pgt{} tries to generalize the
  goal \fancy{T}. Here, \pgt{} alone cannot determine over
  which constant or sub-terms it should generalize~\fancy{T}. 
  Hence, it creates a generalized version of~\fancy{T} for each constant and
  sub-term collected in the previous step. 
  For \verb|[]| in the running example, 
  \pgt{} creates the following generalized version of~\fancy{T}:
  \texttt{!!Nil. itrev xs Nil = rev xs}.

\begin{figure}[t]
    \centering \scalebox{0.5}{
      \begin{tikzpicture}[%
        ->, shorten >=2pt, >=stealth, node distance=0.4cm,
        noname/.style={%
          ellipse, minimum width=5em, minimum height=3em, draw }
        ,scale=0.7]
        \tikzstyle{every node}=[font=\LARGE] \node[rectangle,draw] (1)
        {Input: the original goal \fancy{T} and generalized versions of \fancy{T}
         ($= \fancy{C}_0,\dots,\fancy{C}_n$)};
        \node[rectangle,draw] (2) [below=of 1] 
          {Extract constants in \fancy{T} and $\fancy{C}_0,\dots,\fancy{C}_n$}; 
        \node[rectangle,draw] (3) [below=of 2] 
          {For each constant extracted above, 
          find related constants from the corresponding simp rules};
        \node[rectangle,draw] (4) [below=of 3] 
          {Traverse generalized conjectures and mutate their sub-terms in a
          top-down manner}; 
        \path (1) edge node {} (2) (2) 
                  edge node {} (3) (3) 
                  edge node {} (4) (4);
      \end{tikzpicture}
    }
    \caption{The workflow of the \texttt{conjecture} function.}
    \label{fig-simpConj}
  \end{figure}
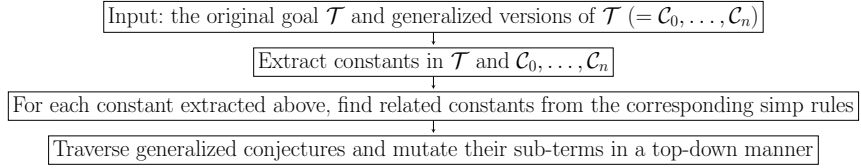

  \paragraph{Goal Oriented Conjecturing.}
  This step calls the function \verb|conjecture|,
  illustrated in Fig.~\ref{fig-simpConj},
  with the original goal \fancy{T} and
  each of the generalized versions of \fancy{T} %
  from the previous step 
  ($\fancy{C}_0,\dots,\fancy{C}_n$). 
  The following code snippet shows part of \verb|conjecture|:
  \begin{verbatim}
  fun cnjcts t = flat (map (get_cnjct generalisedT t) consts)
  fun conj (trm as Abs (_,_,subtrm)) = cnjcts trm @ conj subtrm
   |  conj (trm as App (t1,t2)) = cnjcts trm @ conj t1 @ conj t2
   |  conj trm = cnjcts trm
  \end{verbatim}
  
  For each \fancy{T} and $\fancy{C}_i$ for $0 \leq i \leq n$,
  \verb|conjecture| first calls \verb|conj|,
  which traverses the term structure of each \fancy{T} or $\fancy{C}_i$
  in a top-down manner.
  In the running example,
  \pgt{} takes some $\fancy{C}_k$, say \texttt{!!Nil. itrev xs Nil = rev xs}, as an input and applies \verb|conj| to it.
  
  For each sub-term the function 
  \texttt{get\_cnjct} in \verb|cnjcts| creates new conjectures 
  by replacing the sub-term (\verb|t| in \verb|cnjcts|) 
  in \fancy{T} or 
  $\fancy{C}_i$ (\verb|generalisedT|)
  with a new term. This term is generated from the sub-term (\verb|t|) 
  and the constants (\verb|consts|).
  These are obtained from simplification rules 
  that are automatically derived from the definition of a constant that
  appears in the corresponding \fancy{T} or $\fancy{C}_i$.
  
  In the example, 
  \pgt{} first finds the constant \verb|rev| within $\fancy{C}_k$.
  Then, \pgt{} finds the simp-rule (\texttt{rev.simps(2)}) relevant to 
  \verb|rev|
  which states, \texttt{rev (?x \# ?xs) = rev ?xs @ [?x]}, in the background context.
  Since \texttt{rev.simps(2)} uses the constant \verb|@|, 
  \pgt{} attempts to create new sub-terms
  using \verb|@| while traversing in the syntax tree of
  \texttt{!!Nil. itrev xs Nil = rev xs} in a top-down manner.
  
  When \verb|conj| reaches the sub-term \texttt{rev xs},
  \texttt{get\_cnjct} creates new sub-terms using this sub-term, 
  \verb|@| (an element in \verb|consts|), and
  the universally quantified variable \verb|Nil|.
  One of these new sub-terms would be \texttt{rev xs @ Nil}\footnote{Note that 
  \texttt{Nil} is a universally quantified variable here.}.
  Finally, \texttt{get\_cnjct} replaces the original sub-term \verb|rev xs| with
  this new sub-term in $\fancy{C}_k$,
  producing the conjecture: \texttt{!!Nil. itrev xs Nil = rev xs @ Nil}.
  
\begin{figure}[t]
    \center
    \includegraphics[width=0.5\textwidth]{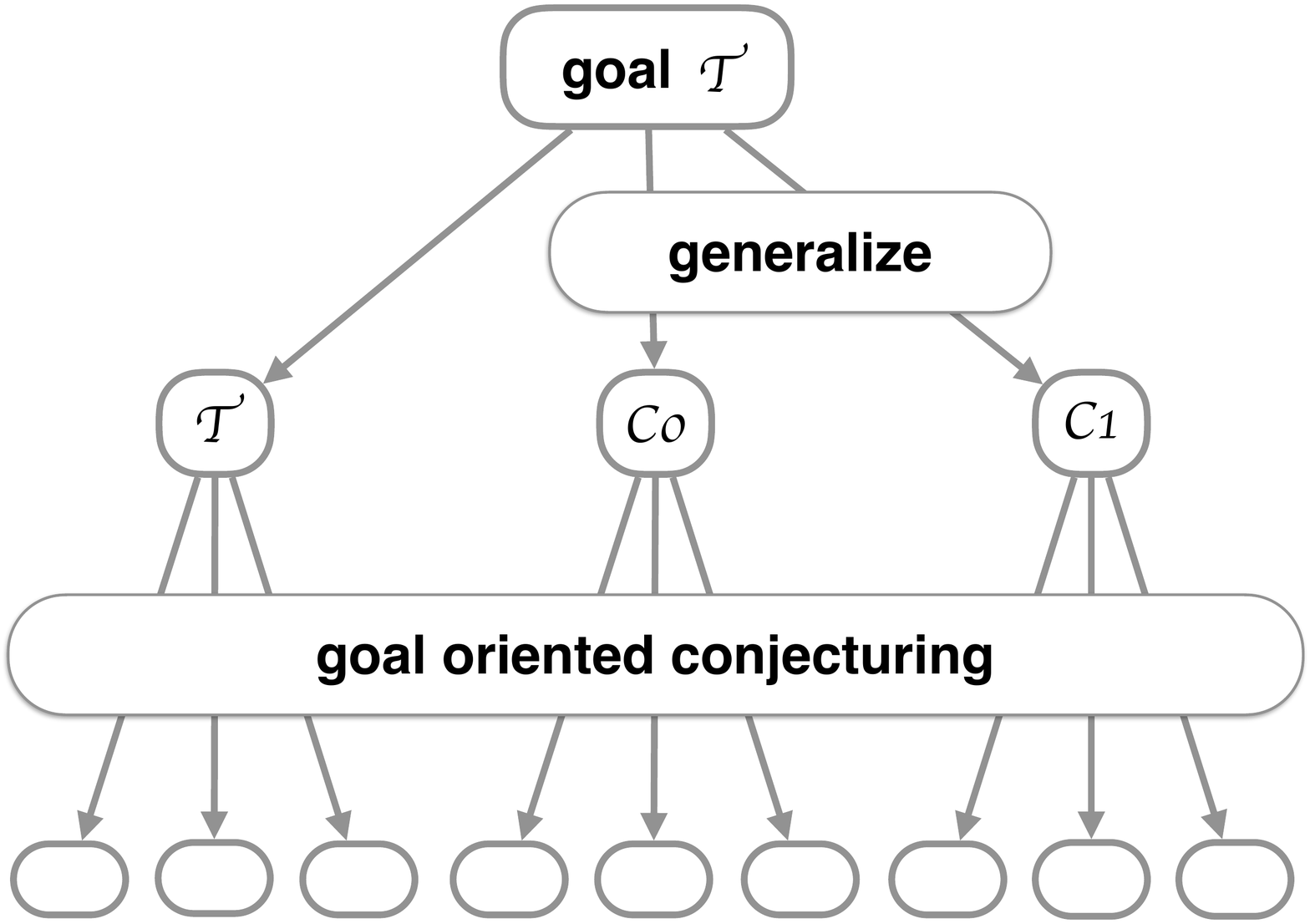}
    \caption{\psl{}'s sequential generalization and goal oriented conjecturing.}
    \label{fig:goal_oriented_conjecturing}
  \end{figure}

  Note that this conjecture is not the only conjecture produced in this step:
  \pgt{}, for example, also produces \texttt{!!Nil. itrev xs Nil = Nil @ rev xs},
  by replacing \verb|rev xs| with \verb|Nil @ rev xs|,
  even though this conjecture is a non-theorem.
  Fig. \ref{fig:goal_oriented_conjecturing} illustrates the sequential application
  of generalization in the previous paragraph and goal oriented conjecturing
  described in this paragraph.
  
  \paragraph{Clean \& Return}
  Most produced conjectures do not even type check.
  This step removes them as well as duplicates
  before passing the results to the following sub-strategy 
  (\texttt{Then [Fastforce, Quickcheck, DInd]} 
  in the example).
  %
  %
  %
  %
  %
  %
  \section{Conclusion}
  We presented an automatic conjecturing tool \pgt{} and
  its integration into \psl{}. 
  Currently, \pgt{} tries to generate conjectures using previously derived simplification rules as hints.
  We plan to include more heuristics to 
  prioritize conjectures before passing them to subsequent strategies.
  
  Most conjecturing tools for Isabelle,
  such as \textit{IsaCoSy}
  \cite{isacosy} and \textit{Hipster} \cite{hipster},
  are based on the bottom-up approach called
  \textit{theory exploration} \cite{theoryexploration}.
  The drawback is that they tend to produce uninteresting conjectures.
  In the case of IsaCoSy the user is tasked with pruning these by hand.
  Hipster uses the difficulty of a conjecture's proof
  to determine or measure its usefulness.
  Contrary to their approach, \pgt{} produces conjectures
  by mutating original goals.
  Even though \pgt{} also produces unusable conjectures internally,
  the integration with \psl{}'s search framework
  ensures that \pgt{} only presents conjectures
  that are indeed useful in proving the original goal.
  Unlike Hipster, which is based on a Haskell code base,
  \pgt{} and \psl{} are an Isabelle theory file,
  which can easily be imported to any Isabelle theory.
  Finally, unlike Hipster, \pgt{} is not limited to equational conjectures.
  
  Gauthier \textit{et al.} described conjecturing across proof corpora \cite{tgck-lpar15}. 
  While \verb|PGT| creates conjectures by mutating the original goal,
  Gauthier \textit{et al.} produced conjectures 
  by using statistical analogies extracted from large formal libraries \cite{conjecture}.
  %
  %
  \bibliographystyle{splncs04} \bibliography{biblio}
  \newpage
  \section*{Appendix}
  \centering
  \begin{figure*}[!ht]
        \centerline{\includegraphics[width=120mm]{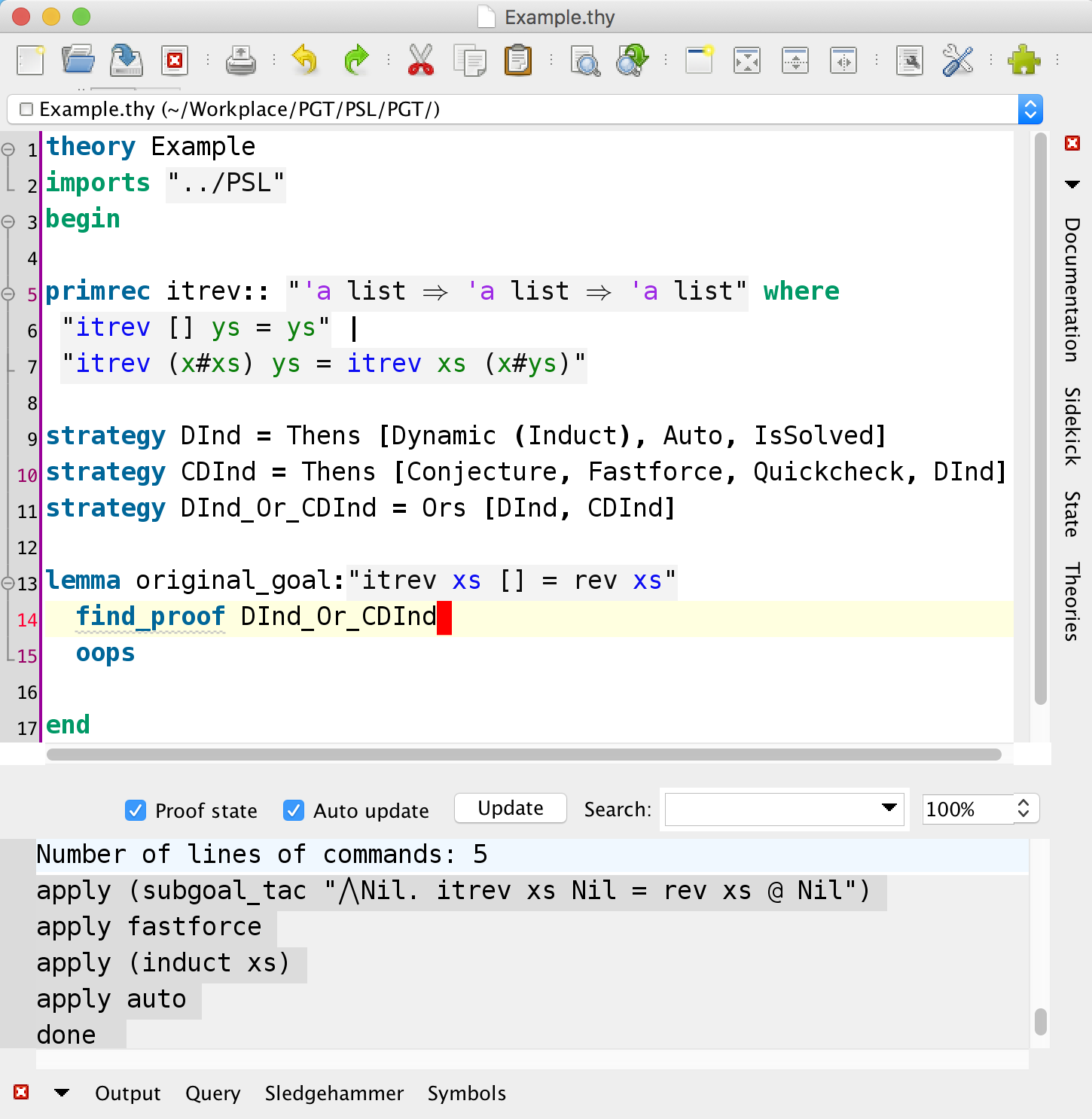}}
        \captionsetup{labelformat=empty}
        \caption{Screenshot of Isabelle/HOL with \pgt{}.}
        \label{fig:screenshot}
  \end{figure*}
  \end{document}